\documentclass[12pt]{article}
\usepackage{fancyhdr}

\usepackage{overcite}

\usepackage{graphicx}

\usepackage{amsfonts}

\usepackage{amsmath}

\usepackage{siunitx}

\usepackage{epstopdf}

\usepackage{caption}

\usepackage{subcaption}

\usepackage{authblk}

\usepackage{setspace}

\usepackage{multirow}

\usepackage{makecell}

\makeatletter \renewcommand\@biblabel[1]{$^{#1}$} \makeatother
\newlength{\bibhang}
\title{Technical Note: Taking EGSnrc to new lows: Development of egs++ lattice geometry and testing with microscopic geometries}
\author{Martin P. Martinov}
\author{Rowan M. Thomson}

\newcommand{\mfig}[1]{\marginpar{{\sf Fig~\ref{#1} }}}
\newcommand{\mtab}[1]{\marginpar{{\sf Table~\ref{#1} }}}

\newcommand{\ie}{{\it i.e.}, }
\newcommand{\eg}{{\it e.g.}, }
\newcommand{\etal}{{\it et al}}

\newcommand{\cen}[1]{\begin{center} #1 \end{center}}

\newcommand{\captionl}[2]{\parbox{16.5cm}{\caption[#1]{{\sf #2}}}}

\newcommand{\um}{\si{\micro\meter}{} }
\newcommand{\nm}{\si{\nano\meter}{} }
\newcommand{\mm}{\si{\milli\meter}{} }

\newcommand{\ums}{\si{\micro\meter}{}}
\newcommand{\nms}{\si{\nano\meter}{}}

\newcommand{\cms}{\si{\centi\meter}{}}

\begin{document}

\cen{\sf {\Large {\bfseries Technical Note: Taking EGSnrc to new lows: Development of egs++ lattice geometry and testing with microscopic geometries} \\
Martin P. Martinov$^\text{a)}$ and Rowan M. Thomson$^\text{b)}$} \\ \vspace{3mm}
email: a) martinov@physics.carleton.ca and b) rthomson@physics.carleton.ca \\ \vspace{5mm}
Carleton Laboratory for Radiotherapy Physics, Department of Physics, Carleton University, Ottawa, Ontario, K1S 5B6, Canada \\ \vspace{5mm} 
}
\pagestyle{empty}

%

\begin{abstract}

\noindent\textbf{Purpose:} This work introduces a new lattice geometry library, egs\_lattice, into the EGSnrc Monte Carlo code, which can be used for both modeling very large (previously unfeasible) quantities of geometries (\eg cells or gold nanoparticles) and establishing recursive boundary conditions.  The reliability of egs\_lattice, as well as EGSnrc in general, is cross-validated and tested at short length scales and low energies.

\noindent\textbf{Methods:} New Bravais, cubic, and hexagonal lattice geometries are defined in egs\_lattice and their transport algorithms are described.  Simulations of cells and Gold NanoParticle (GNP) containing cavities are implemented to compare to independent, published Geant4-DNA and PENELOPE results.  Recursive boundary conditions, implemented through a cubic lattice, are used to perform electron Fano cavity tests.  The Fano test is performed on three different-sized cells containing GNPs in the region around the nucleus for three source energies. 
  
\noindent\textbf{Results:} Lattices are successfully implemented in EGSnrc, and are used for validation.  EGSnrc calculated the dose to cell cytoplasm and nucleus when irradiated by an internal electron source with a median difference of 0.6\% compared to published results Geant4-DNA results.  EGSnrc calculated the ratio of dose to a microscopic cavity containing GNPs over dose to a cavity containing a homogeneous mixture of gold, and results generally agree (within 1\%) with published PENELOPE results.  The electron Fano cavity test is passed for all energies and cells considered, with sub-0.1\% discrepancies between EGSnrc-calculated and expected values.  Additionally, the recursive boundary conditions used for the Fano test provided a factor of over a million increase in efficiency in some cases.  

\noindent\textbf{Conclusions:} The egs\_lattice geometry library, currently available as a pull request on the EGSnrc GitHub ``develop'' branch, is now freely accessible as open-source code.  Lattice geometry implementations cross-validated with independent simulations in other MC codes and verified with the electron Fano cavity test demonstrate not only the reliability of egs\_lattice, but, by extension, EGSnrc's ability to simulate transport in nanometer geometries and score in microscopic cavities.

\end{abstract}

\newpage
\setlength{\baselineskip}{0.7cm}
\pagestyle{fancy}

\section{Introduction} \label{sec_EGSnrc:Introduction} 

Monte Carlo (MC) simulations involving many repeated geometric elements can present heavy computational burdens.  Excessive memory requirements for storing information about each (repeated) geometric element may prevent execution of simulations.  Further, boundary checking requirements during particle transport may lead to prohibitively long calculation times.  Lattice geometries provide an alternate modeling approach that, depending on implementation, may enable less memory-intensive and time-consuming calculations.  Recent reports involving lattice geometrics in MC simulations include discretely-modeled GNPs in a large volume using a cubic lattice of GNPs embedded in an otherwise homogeneous water or tissue phantom\cite{Me13,Gh12,Gh13,Zh09,To12b} for various scenarios.  Cai \etal\cite{Ca13} have implemented a hexagonal lattice of GNPs in a cell.  In general, the description of the implementation of the lattice within particular MC codes is limited and varies between publications.

This technical note introduces egs\_lattice, a new egs++ geometry library that allows for modeling of Bravais, cubic, and hexagonal lattices in the EGSnrc MC code.  We describe in detail the algorithms behind the cubic lattice geometry used in previous work\cite{MT17}, as well as a more general Bravais lattice and a more complex hexagonal close-packed lattice.  The egs\_lattice geometry library is currently freely-available as a pull request on the EGSnrc GitHub distribution (``develop'' branch).  The lattices can be used for modeling of infinitely-repeating geometries or establishing recursive boundary conditions.  The lattice geometries allow for modeling of very large numbers of geometries (such as cells or gold nanoparticles in a tumor), significantly increasing EGSnrc's performance when modeling systems with very short length scale structures.

The EGSnrc MC code has been a gold standard for many medical physics applications, such as metrology\cite{RK03}, though is mostly used in the context of MV energy irradiations looking at voxelized centimeter and millimeter-scale geometries.  As such, much of the EGSnrc testing and validation has been performed for patient-scale phantoms or is specific to ion chamber simulations\cite{Ka99b,Wa00,AM15,MR16}.  EGSnrc has been used at short length scales as well, such as cell dosimetry\cite{Th13,OT18,OT19}, nanofilm spectrometry\cite{AR08}, and microdosimetric studies\cite{Ve05,Sy04,OT18a}.  However, the validation of EGSnrc in these works is typically focused on the specific scenarios considered.  This technical note, introducing the egs\_lattice geometry library, provides an opportunity to validate not only the new egs\_lattice geometry, but EGSnrc's ability to score and perform transport on short length scales on a more general level.

Several different scenarios are simulated to cross-validate and test EGSnrc, as well as the egs\_lattice geometry.  Cellular dosimetry in a Medical Internal Radiation Dosimetry scenario is performed and compared with published Geant4-DNA results\cite{Se15b}:  S-values are computed for four different source/target configurations.  The ratio of dose to the tissue in the presence of a lattice of GNPs divided by dose to an equivalent gold-tissue mixture is computed and compared with published PENELOPE results\cite{KK16}.  Ratios are computed for three different energies, gold concentrations and GNP sizes.

EGSnrc is further verified with the electron Fano cavity test\cite{Bo15}.  The Fano test, based on Fano's theorem\cite{Fa54a}, is considered fundamental to validate MC codes of radiation transport, as it sets up a scenario in which dose calculations can be benchmarked against an analytical dose expression.  Thus, the Fano cavity test provides an additional check on EGSnrc's ability to model transport through complex geometries, independent of other MC work.  A cell containing a hexagonal lattice of GNPs near its nucleus is placed in a box with recursive boundaries (formed using cubic lattice) is tested to ensure the calculated dose is within 0.1\% of the predicted dose, a threshold for EGSnrc established previously on ionization chambers\cite{Ka99b,Se02}.

\section{Methods} \label{sec_EGSnrc:Methods}

MC simulations are carried out using EGSnrc\cite{Ka11} with the egspp class library\cite{Ka05a}, pulled from commit 822ec3a of the EGSnrc GitHub develop branch.  A modified version of egs\_chamber\cite{Wu08}, with the ability to score energy-deposition in multiple regions, is used for all simulations.

Default transport parameters for EGSnrc are used with the following exceptions: pair angular sampling is turned off, Rayleigh scattering is turned on, and electron impact ionization is turned on (using the default Electron Impact Ionization, EII, option).  Bremsstrahlung events and photon interactions use NRC and XCOM cross section data\cite{Be10}, respectively.  In addition to the $K$- and $L$-shells explicitly, the compiler option to model $M$- and $N$-shell transitions explicitly as well is used.  Particle transport is modeled down to 1~keV kinetic energy.  All media used in the MC simulations are either gold, water, ICRU four-component tissue (10.1\% hydrogen, 11.1\% carbon, 2.6\% nitrogen, and 76.2\% oxygen by mass\cite{ICRU44}), or a homogeneous blend of gold with tissue or water.

\subsection{Lattice geometry implementation in EGSnrc} \label{ssec:Methods-latt}

Traditionally, EGSnrc only comes with a repeater-type geometry class library\cite{Ka05a}, which is difficult to use and has many limitations (especially when repeating in three dimensions).  A lattice geometry class library is created to allow for modeling of Bravais, cubic, or hexagonal close-packed lattices that overcome many of the repeater class limitations.  The lattice geometries only store two unique geometries in memory, the geometry in which the lattice is embedded and the geometry defined at each lattice position; this modeling approach is limited in that one must score in all geometries in the lattice at once (or none), but the approach can accommodate an arbitrary number of geometries, \eg{} handling the modeling of trillions of geometries as ably as ten.  This subsection describes the implementations of the egs\_lattice geometry in EGSnrc.

The lattice geometry takes a predefined geometry, referred to as the `subgeometry', and places it at points defined by a Bravais lattice\cite{Ar06a} within a chosen region of another geometry, referred to as the `holder'.  A vector representing a point (at which the center of the subgeometry would be placed) along a Bravais lattice is defined as
\begin{equation}
		\vec{r}(i,j,k) = i a_x \hat{x} + j a_y \hat{y} + k a_z \hat{z} \label{eq:latt}
\end{equation}
$\text{where } a_x, a_y, a_z \in {\mathbb R}\text{ and } i, j, k \in {\mathbb Z}$.  The same instance of the subgeometry is placed at all possible positions $\vec{r}$ within a region of the holder, \ie{} dose to all subgeometries is scored at once.  The lattice geometries provide substantial reductions in memory use (only one subgeometry is stored in memory) and radiation transport time (only transport through subgeometries positioned near the photon or electron path) compared to a simulation in which each subgeometry would be modeled explicitly.

\begin{figure}[htbp]
	\centering
	\includegraphics[width=0.45\textwidth]{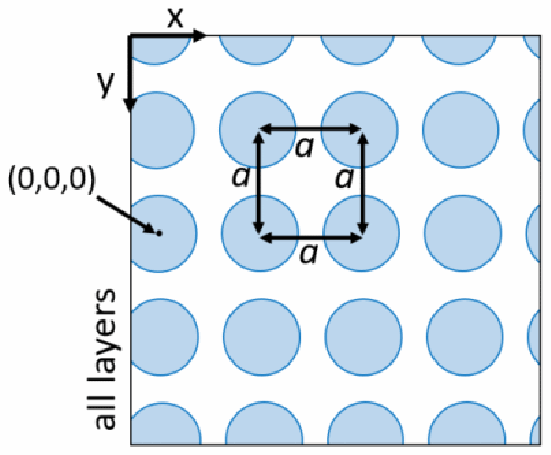}
	\captionl{Cubic lattice diagram}{2D cross-sectional diagram showing an arbitrary slice of a cubic lattice layer of spheres with center-to-center distance $a$.
	\label{fig:CubeLatt}}
\end{figure}

\begin{figure}[htbp]
	\centering
	\includegraphics[width=0.45\textwidth]{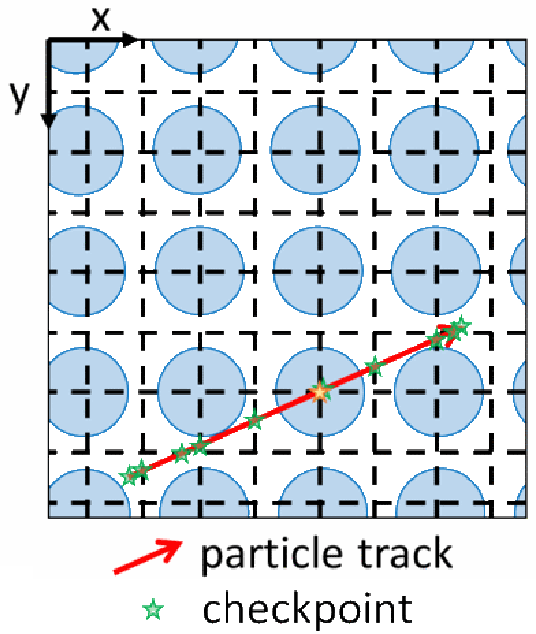}
	\captionl{Cubic lattice transport example}{Diagram depicting a 2D example of cubic lattice \textit{howfar}() algorithm.  The red line is the track for which \textit{howfar}() is invoked and the green stars are the positions (checkpoints) at which the \textit{howfar}() algorithm is called for the nearest subgeometry.  The gold star is the first checkpoint that returns an intersection result from a subgeometry.
	\label{fig:CubeLattTrans}}
\end{figure}

The same subgeometry repeats in each of the $x$, $y$, and $z$ directions with the same spacing $a_x$, $a_y$, and $a_z$ (respectively) around a subgeometry defined at the origin (Fig.~\ref{fig:CubeLatt}\mfig{fig:CubeLatt} shows a cubic lattice with $a_x$, $a_y$, and $a_z$ set to a single value $a$).  In EGSnrc, there are three primary functions for particle transport:
\begin{itemize}
	\item \textit{isWhere}() gets passed a position [$x$,$y$,$z$] and returns the geometry region in which that point lies.  In the case of the lattice, when the point is in a holder region that contains subgeometries, the subgeometry is placed at the nearest lattice position to that point (\ie [$a_x\cdot{}$round$(x/a)$,$a_y\cdot{}$round$(y/a_y)$,$a_z\cdot{}$round$(z/a_z)$]) and the \textit{isWhere}() of the subgeometry is invoked.  If the [$x$,$y$,$z$] is within the subgeometry, then the subgeometry \textit{isWhere}() call is returned otherwise the holder \textit{isWhere}() is returned.
	\item  \textit{hownear}() returns the distance to the closest boundary from a position [$x$,$y$,$z$].  When invoked in the subgeometry or holder geometry regions not containing the lattice, it invokes those geometries' native \textit{hownear}() calls.  Otherwise, the subgeometry is placed at the eight nearest lattice positions to that point (one index up or down from [$x$,$y$,$z$]) which \textit{draw} a rectangular prism enclosing [$x$,$y$,$z$].  The shortest \textit{hownear}() call of the eight subgeometries and the holder is returned.
	\item \textit{howfar}() is a function which, in brief, takes a (particle's) position [$x$,$y$,$z$], direction of motion [$p$,$q$,$r$], and distance it would travel in the medium, and then calculates if the distance fits in the invoked region or how far the particle would travel in the region before leaving.  As in the \textit{hownear}() case, when invoked in the subgeometry or holder geometry regions not containing the lattice, it invokes those geometries' native \textit{howfar}() calls.  If the call is in the holder region with subgeometries, the algorithm draws the track from the start position until it travels the full distance or leaves the holder region.  At every point that the track intersects with planes defined at $x=n\cdot{}\frac{a_x}{2}$, $y=n\cdot{}\frac{a_y}{2}$, and $z=n\cdot{}\frac{a_z}{2}$ as well as the starting and ending track positions, a point on the track is marked (Fig.~\ref{fig:CubeLattTrans}\mfig{fig:CubeLattTrans}).  For each marked point along the track (ordered from start to end), the subgeometry is placed at the nearest lattice position to that point and \textit{howfar}() is invoked for that subgeometry.  The algorithm stops iterating if there is an intersection with a subgeometry or returns the \textit{howfar}() of the holder if no intersections are found.
\end{itemize}

This boundary crossing algorithm is exact if the subgeometries are centered at their lattice positions and if any radial asymmetries in the subgeometry boundary are not larger than a quarter of the center-to-center distance between subgeometries.  Any implementation in this work only uses subgeometries with spherically symmetric boundaries.

\begin{figure}[htbp]
	\centering
	\includegraphics[width=0.45\textwidth]{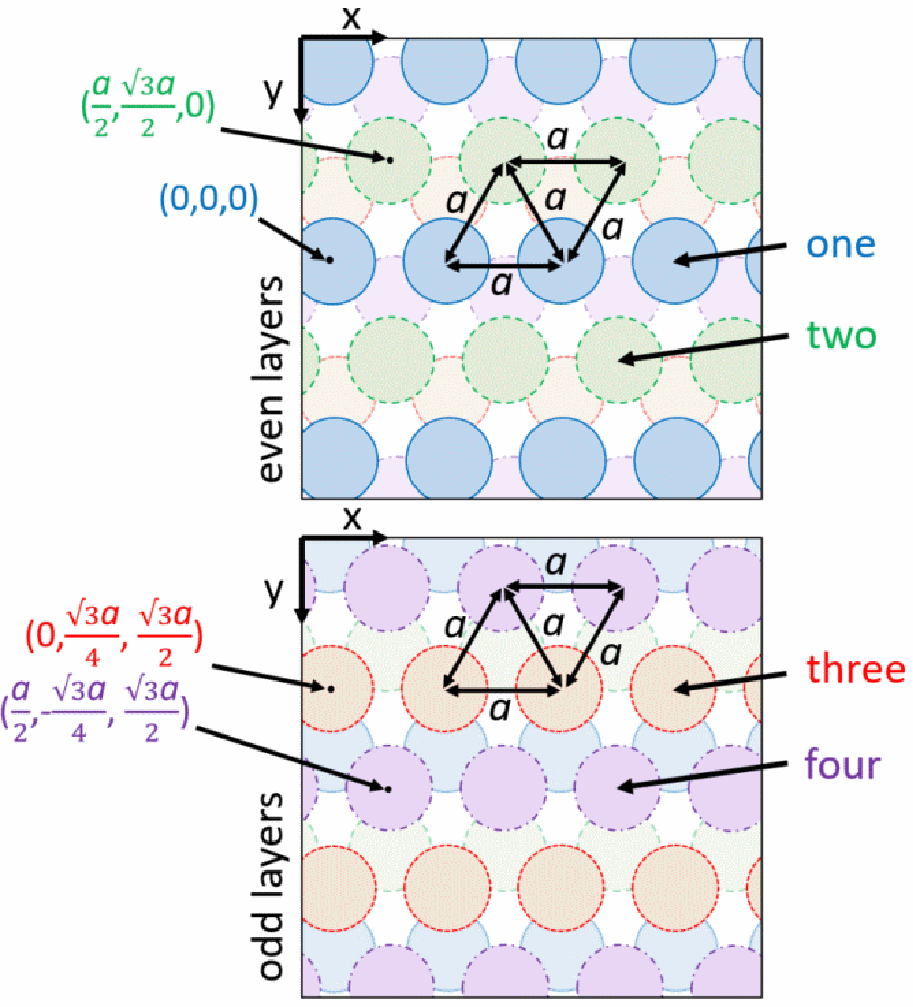}
	\captionl{Hexagonal lattice diagram}{2D cross-sectional diagram showing an arbitrary slice of two hexagonal lattice layers of spheres with center-to-center distance $a$.  The four different Bravais lattices defined by equation~(\ref{eq:hex_latt}) are shaded/outlined separately for clarity.
	\label{fig:HexLatt}}
\end{figure}

Using much of the above Bravais lattice architecture, a hexagonal close-packed lattice is developed using four overlapping lattices defined above (Fig.~\ref{fig:HexLatt}\mfig{fig:HexLatt}) to achieve a higher packing density of subgeometries.  The transport algorithms for the hexagonal lattice is an amalgamation of transport through one Bravais lattice ($\vec{r_1}$) and three of its translations ($\vec{r_2}$, $\vec{r_3}$, and $\vec{r_4}$), defined as
\begin{equation} \label{eq:hex_latt} 
	\centering
	\begin{array}{rcrrcrrcrr}
		\vec{r_1}(i,j,k) &=&               i&a \hat{x} &+&               j&\sqrt{3}a \hat{y} &+&               k&\sqrt{3}a \hat{z} \\
		\vec{r_2}(i,j,k) &=& (i+\frac{1}{2})&a \hat{x} &+& (j+\frac{1}{2})&\sqrt{3}a \hat{y} &+&               k&\sqrt{3}a \hat{z} \\
		\vec{r_3}(i,j,k) &=&               i&a \hat{x} &+& (j+\frac{1}{4})&\sqrt{3}a \hat{y} &+& (k+\frac{1}{2})&\sqrt{3}a \hat{z} \\
		\vec{r_4}(i,j,k) &=& (i+\frac{1}{2})&a \hat{x} &+& (j-\frac{1}{4})&\sqrt{3}a \hat{y} &+& (k+\frac{1}{2})&\sqrt{3}a \hat{z}
	\end{array}
\end{equation}
${\text{where } a \in {\mathbb R}\text{ and } i, j, k \in {\mathbb Z}}$, with changed \textit{isWhere{}}, \textit{howfar}(), and \textit{hownear}() implementations which check against all four lattices at once to improve efficiency.

\subsection{Cross-validation of EGSnrc} \label{ssec:Methods-Valid}

\subsubsection{Scoring dose in cellular geometries}

\begin{figure}[htbp]
	\centering
	\includegraphics[width=0.45\textwidth]{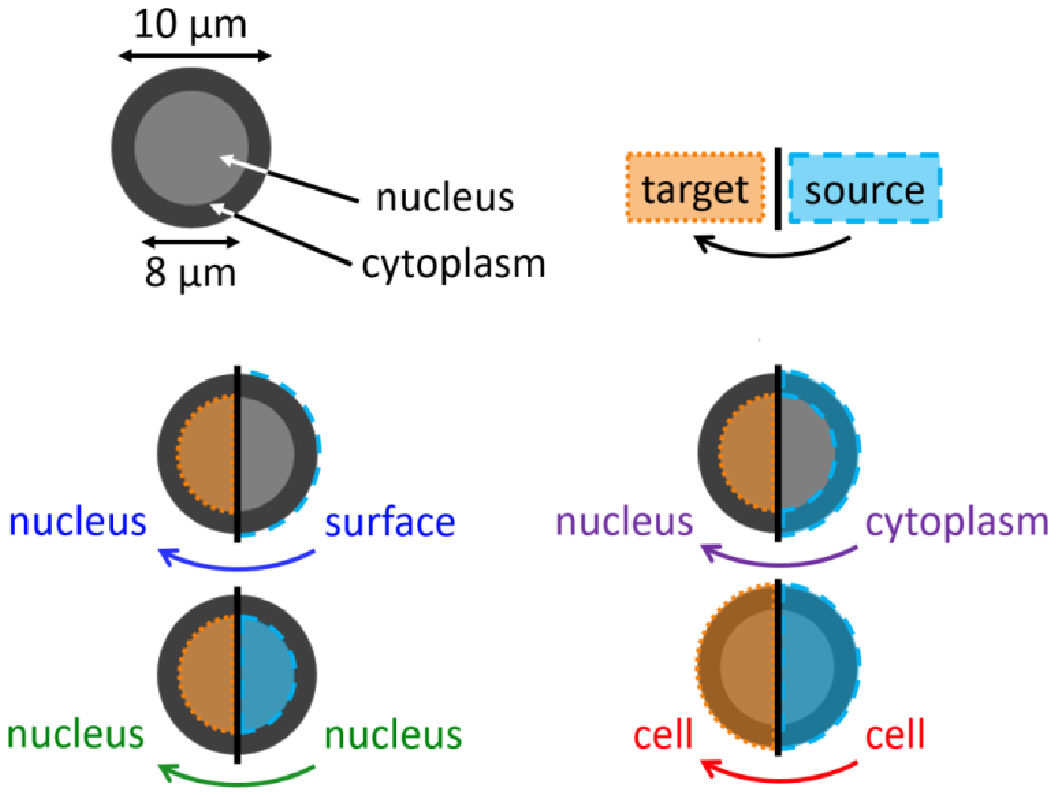}
	\captionl{S-value calculation diagram}{Cross section diagram of the cell used for the S-value calculations (top) and the four scenarios (labeled as target$\leftarrow$source) compared to Geant4-DNA.
	\label{fig:SeflDiagram}}
\end{figure}

EGSnrc's ability to score dose to cellular sub-compartments (\ie nucleus and cytoplasm) is compared to the Geant4-DNA results of {\v{S}}efl \textit{et al}\cite{Se15b}.  Medical Internal Radiation Dose (MIRD) S-value calculations are performed on a cell modeled as two concentric spheres, an inner nucleus and outer cytoplasm, with dimensions ($r_\text{cell}$, $r_\text{nuc}$) = (5, 4)~\ums, where the S-value is the dose absorbed by a target (\ie the nucleus, cytoplasm, or both) per activity per second of a specific source configuration.  Isotropic, monoenergetic electron sources are distributed in four different configurations: on the surface of the cell, throughout the entire cell, throughout only the cytoplasm or throughout only the nucleus.  Doses to the nucleus and cytoplasm are scored for electron energies from 1 to 100~keV and are used to calculate S-values for four different scenarios.  S-values are therefore equivalent to dose/history as the number of histories and electrons are the same.  All cell compartments are modeled as liquid water.  Four source/target scenarios, outlined in Fig.~\ref{fig:SeflDiagram}, are simulated and compared with calculations using Geant4-DNA.  Each scenario is simulated to $4\times10^9$ histories to achieve uncertainties below 0.01\% for all but the near-zero dose cases.

\subsubsection{Simulation of nanometer geometries using a lattice}

In previous work\cite{MT17}, the ratio of dose-to-tissue over dose to the homogeneous blend of tissue and gold were calculated using a cubic lattice and compared to results published by Koger and Kirkby\cite{KK16}, the EGSnrc and lattice geometry validation is continued here.  Simulations where monoenergetic photons (20, 30, and 50~keV) irradiating one face of a cylinder containing GNPs in ICRU 4-component tissue\cite{ICRU44} (concentrations of 5, 10, and 20 mg of gold per g of tissue, shortened to mg/g) are performed.  The cylinder is (150~\um radius, 200~\um long) with a smaller cylinder (100~\um radius, 100~\um long) located at the center of the larger cylinder used for scoring.  This work is a follow-up to validation in one previous work\cite{MT17}, where a more detailed description of these simulations can be found\cite{MT17} (Sec.~II.A and III.A).  In addition to the 20 and 100~\si{\nano\metre} diameter GNP values calculated previously, 50~\si{\nano\metre} GNPs are also investigated.  These simulations are performed using a cubic and a hexagonal close-packed lattice (both with same number of GNPs), to ensure consistency.

\subsection{Electron Fano cavity test}  \label{ssec:Methods-Fano}

\begin{figure}[htbp]
	\centering
	\includegraphics[width=0.6\textwidth]{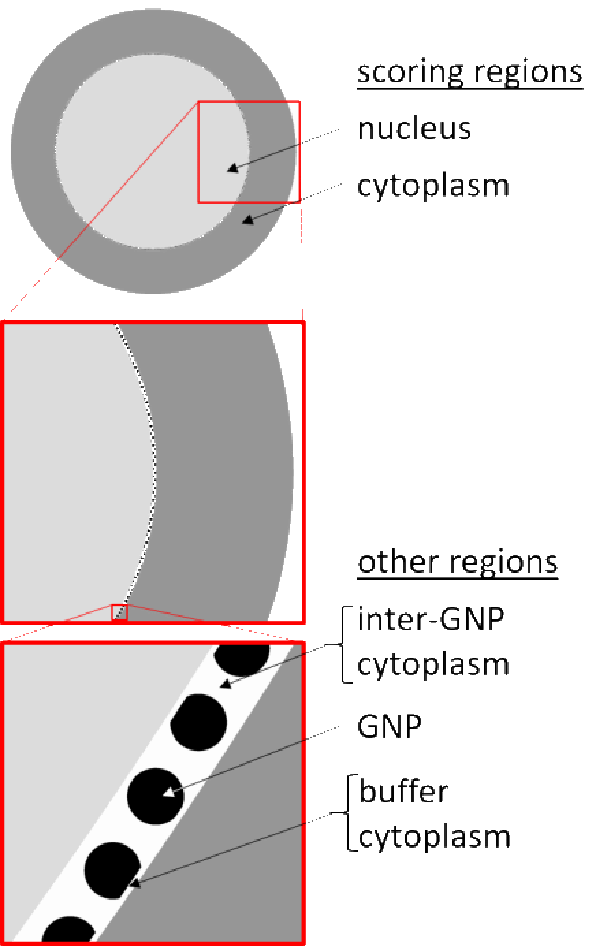}
	\captionl{Electron Fano cavity test image}{EGSnrc-generated images, using application egs\_view, of a ($r_\text{cell}$, $r_\text{nuc}$) = (7.35, 5)~\um for a 20~mg/g gold concentration used in the electron Fano cavity tests.  GNPs have a radius of 25~\nms.
	\label{fig:FanoCell}}
\end{figure}

An electron Fano cavity test, based on work presented by Senpau and Andreo\cite{SA06} and Bouchard \textit{et al}\cite{Bo15} for testing transport in magnetic fields, is performed (with zero magnetic field) to test electron transport in EGSnrc.  Three cells with a ($r_\text{cell}$, $r_\text{nuc}$) of (5, 3)~\ums, (7.35, 5)~\ums, or (10, 8)~\um with a hexagonal lattice of 19624, 52723, or 119826 GNPs (respectively) in a thin shell (Fig.~\ref{fig:FanoCell}\mfig{fig:FanoCell}), a complex microscopic scale model used in other work\cite{MT20}, is used for the test.  The cell is placed in a small cube of tissue (side length 25~\ums), with uniform material elemental composition (ICRU tissue), but non-uniform mass density (\ie GNPs are modeled as spheres of 19.32~g/\si{\centi\metre}$^3$ density tissue with a density effect correction for nominal tissue density).  An isotropic electron source is distributed throughout the cell and cube, generating a uniform number of electrons per unit mass everywhere (\ie 19.32 times as many electrons per unit volume in the GNPs as elsewhere).  Recursive boundary conditions are established; if a particle crosses the x, y, or z boundaries of the cube, it is translated to the opposite boundary while maintaining its velocity.  The recursive boundary condition is fulfilled by using the cell as the subgeometry in a cubic lattice with a center-to-center distance of 25~\ums.

In Charged Particle Equilibrium (CPE) conditions, with the source used in this simulation, the dose per history calculation is very simple, \ie
\begin{equation}
	\begin{aligned}
		\text{dose} &= \frac{\text{electron energy}\cdot\text{number of electrons}}{\text{Fano cube mass}} \\
		\frac{\text{dose}}{\text{history}} &= \frac{\text{electron energy}}{\text{Fano cube mass}} \label{eq:CPE}
	\end{aligned}
\end{equation}
because the number of histories is the number of generated electrons.  Three cell sizes are tested: ($r_\text{cell}$, $r_\text{nuc}$) = (5, 3)~\ums, (7.35, 5)~\ums, and (10, 8)~\ums.  Source energies of 20, 30, 50, 100, and 1000~keV are used and a gold concentration of 20 mg/g (25~\nm GNP radius) is considered.

Simulations of a single cell in infinite tissue with no recursive boundary conditions are also performed to demonstrate the efficiency of using the cubic lattice for the Fano test.  In these simulations, the cubic box in which particles are generated is replaced by a spherical source with a radius of two CSDA ranges (3.4~\um to 8.8~\mm for 10~keV to 1~MeV energies)\cite{Be98a} plus the cell cytoplasm radius to ensure CPE.  Single cell and recursive boundary tests are run for $10^9$ histories in most cases except for $2\times10^9$~histories for the small cell at 30~keV, $4\times10^9$~histories for the small cell at 20~keV, and $10^8$~histories for all cells at 1~MeV.

\section{Results} \label{sec_EGSnrc:Results}

\subsection{Cross-validation}

\subsubsection{Scoring dose in cellular geometries \label{ssec:Results-Svalue}}

\begin{figure}[htbp]
	\centering
	\includegraphics[width=0.75\textwidth]{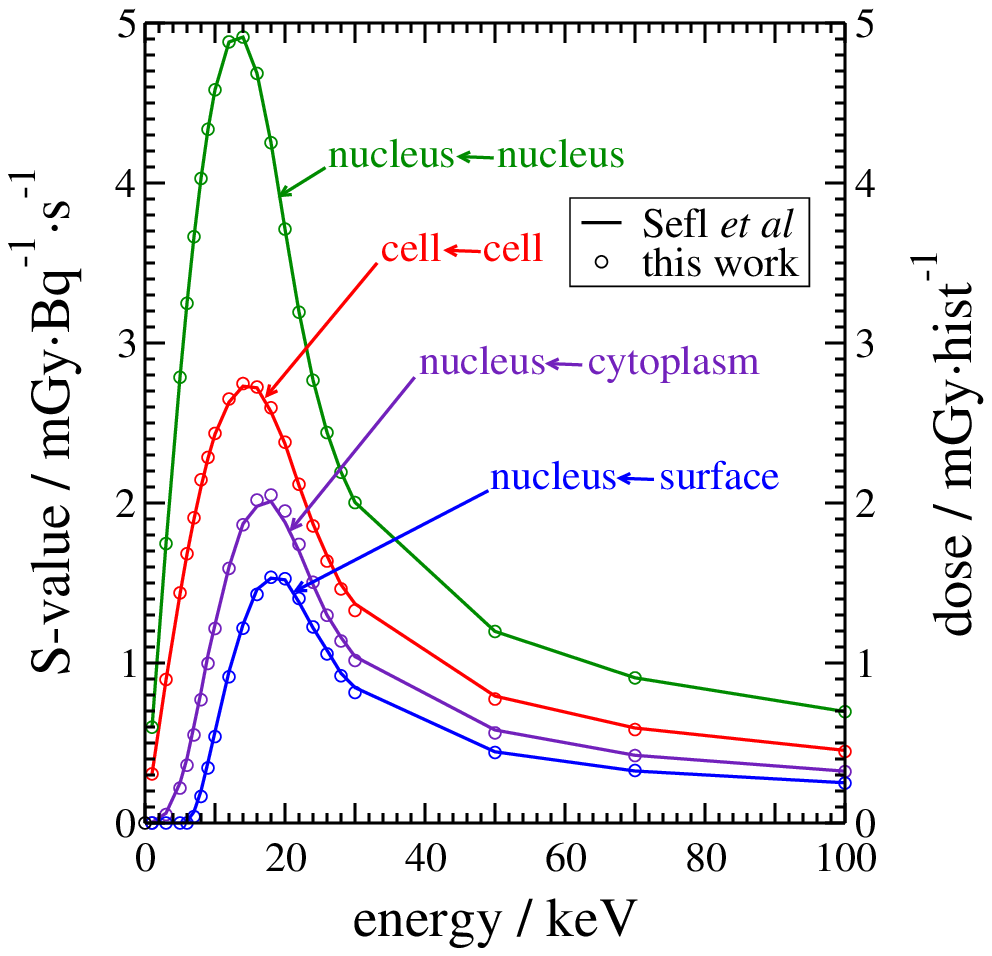}
	\captionl{MIRD S-values}{S-values computed in egs\_chamber compared to values computed by {\v{S}}efl \etal\cite{Se15b} in Geant4-DNA for 4 different scenarios (target$\leftarrow$source).
	\label{fig:SeflPlot}}
\end{figure}

S-values calculations for four different scenarios are shown in Fig.~\ref{fig:SeflPlot}\mfig{fig:SeflPlot} and compared with S-values computed by {\v{S}}efl \etal\cite{Se15b} using Geant4-DNA.  When comparing percent the absolute difference between the Geant4-DNA results with those using egs\_chamber, the majority of S-values agree to sub 1\% levels.  There are a few outliers (up to 3.7\% difference) at 30 and 50~keV for most scenarios, as well as in the 18-22~keV energy range for the nucleus$\leftarrow$cytoplasm scenario.  Additionally, there are some larger percent differences at sub-10~keV energies when the source and target are not congruent for which S-values are low, \eg at 6~keV our S-value (nucleus$\leftarrow$cytoplasm) is 0.361 which is 12\% lower than the 0.404 value of {\v{S}}efl \etal.  Overall, the percent absolute difference is 1.6\% on average with a median difference of 0.57\%, or 1.0\% with a median of 0.40\% when omitting nucleus$\leftarrow$cytoplasm results.  All statistics are calculated excluding sub-0.1~mGy/Bq/s S-values as the doses are very low with high uncertainty.


\subsubsection{Simulation of nanometer geometries using a lattice}

Table~\ref{tab:Koger}\mtab{tab:Koger} presents dose ratios (dose-to-tissue relative to dose to a homogeneous tissue-gold mixture) for various gold concentrations and source energies using a cubic lattice model for GNPs.  While many values computed in the current work agree with the published values of Koger and Kirkby \cite{KK16} within the 1$\sigma$ statistical uncertainties indicated, most values agree within 2$\sigma$ uncertainties.  There is a notable outlier in the 20~mg/g concentration of 50~\nm GNPs at 50~keV conversion factors.  It is of note that the Koger and Kirkby result in this case is much closer to unity than in the 10 or 5~mg/g cases, which opposes the expected trend (more gold leads to larger discrepancy in models, which leads to conversion factors farther from unity) and is likely a statistical outlier.

Omitting the aforementioned data point, the largest percent difference observed is 2.5\% for the 50~keV beam with 20~nm diameter GNPs having a concentration of 20~mg/g.  The absolute difference between our results and those of Koger and Kirkby averaged over all source energies, GNP diameters, and gold concentrations (except 20~mg/g, 50~\nm and 50~keV case) considered is 0.98\%. Results generally agree within statistical uncertainties, with the average absolute difference divided by uncertainty being 0.99.  Simulations repeated with larger cylindrical phantoms (radii of 200 or 300~\um rather than 150~\ums, lengths of 300 or 400~\um rather than 200~\ums) but with the same central scoring volume (radius 100~\ums, length 100~\ums) yield dose ratios in agreement within statistical uncertainties.  The above statistics are all obtained using the cubic lattice results, but for all scenarios investigated, the hexagonal lattice agreed within statistical uncertainty.  Thus, EGSnrc results computed using the cubic and hexagonal lattice are consistent, and results are in agreement with Koger and Kirkby's PENELOPE results \cite{KK16}.

\begin{table}[htbp]
	\centering
	\vspace{0.5\normalbaselineskip}
	\captionl{Homogeneous to discrete GNP conversion factors}{Comparison of dose ratios (dose-to-tissue relative to dose-to-mixture of tissue and gold, \ie $\frac{D_\text{tissue}}{D_\text{mixture}}$) for 20, 50 and 100~\nm diameter GNPs for the current work and the independent results of Koger and Kirkby calculated using PENELOPE\cite{KK16}.  Statistical uncertainty (1$\sigma$) on the final digit(s) is indicated in parentheses, \eg 0.949 (6) means 0.949 $\pm$ 0.006
	\label{tab:Koger}}
	\small
	\begin{tabular}{ccllllllll}
		\hline \hline \vspace{-0.8\normalbaselineskip} \\
		\multirow{2}{*}{ \makecell{GNP\\conc.\\$\frac{\text{mg}_\text{Au}}{\text{g}_\text{tissue}}$} } & \multirow{2}{*}{ \makecell{en.\\keV} } & \multicolumn{2}{c}{ 20 \nm diam. GNPs } & & \multicolumn{2}{c}{ 50 \nm diam. GNPs } & & \multicolumn{2}{c}{ 100 \nm diam. GNPs } \\ \cline{3-4} \cline{6-7} \cline{9-10} \vspace{-0.8\normalbaselineskip} \\
		& & \makecell{Koger and\\Kirkby} & This work & & \makecell{Koger and\\Kirkby} & This work & & \makecell{Koger and\\Kirkby} & This work \\ \hline \vspace{-0.8\normalbaselineskip} \\
		\multirow{3}{*}{ 5 } & 20 & 0.949 (6) & 0.951 (3) & & 0.920 (7) & 0.927 (1) & & 0.951 (3) & 0.927 (1) \\
		& 30 & 0.969 (8) & 0.963 (5) & & 0.945 (7) & 0.948 (3) & & 0.963 (5) & 0.948 (3) \\
		& 50 & 0.984 (12) & 0.990 (9) & & 0.967 (10) & 0.958 (4) & & 0.990 (9) & 0.958 (4) \\ \vspace{-0.6\normalbaselineskip} \\
		\multirow{3}{*}{ 10 } & 20 & 0.937 (6) & 0.926 (2) & & 0.887 (7) & 0.891 (1) & & 0.926 (2) & 0.891 (1) \\
		& 30 & 0.950 (8) & 0.947 (4) & & 0.929 (7) & 0.928 (2) & & 0.947 (4) & 0.928 (2) \\
		& 50 & 0.984 (11) & 0.968 (8) & & 0.960 (10) & 0.941 (4) & & 0.968 (8) & 0.941 (4) \\ \vspace{-0.6\normalbaselineskip} \\
		\multirow{3}{*}{ 20 } & 20 & 0.917 (7) & 0.902 (2) & & 0.866 (7) & 0.857 (1) & & 0.902 (2) & 0.857 (1) \\
		& 30 & 0.952 (8) & 0.939 (3) & & 0.920 (8) & 0.906 (2) & & 0.939 (3) & 0.906 (2) \\
		& 50 & 0.986 (11) & 0.961 (6) & & 0.971 (10) & 0.920 (3) & & 0.961 (6) & 0.920 (3) \\ \hline \hline \vspace{-0.8\normalbaselineskip} \\
	\end{tabular}
	\vspace{0.5\normalbaselineskip}
\end{table}
\newpage

\subsection{Electron Fano cavity test \label{ssec:Methods-Verify}} 

Results of the Fano test are compiled in Table~\ref{fig:FanoTable}\mtab{fig:FanoTable}.  For all cell sizes and energies investigated, with dose/history uncertainties computed to 0.01\% or less, are within 0.1\% of the expected value calculated with equation~(\ref{eq:CPE}).  The calculated value minus the theoretical value is systematically positive.

Single cell simulations with no recursive boundaries also pass the Fano test at 20 and 30~keV energies.  Results at higher energies agreed with the theoretical values but the uncertainties on nucleus and cytoplasm doses for the 50, 100, and 1000~keV simulations are well above the 0.01\% threshold used above.  The efficiency ratios when using recursive boundary conditions over the single cells with no recursive boundaries are compiled in Table~\ref{tab:Eff}\mtab{tab:Eff}.  Efficiency gains when using recursive boundary conditions ranged from a factor of 2.6 (small cell, 20~keV) to 2.5$\times10^6$ (large cell, 1~MeV) increase.

\begin{table}[htbp]
	\centering
	\rotatebox{90}{
	\captionl{Electron Fano cavity test results}{Electron Fano cavity test results of egs\_chamber dose calculations (0.01\% statistical uncertainty) and comparison to expected dose, $\%\Delta = \frac{\text{result}-\text{expected}}{\text{expected}}\cdot100\%$  
	\label{fig:FanoTable}}}
	\raisebox{.35\textheight}[0pt][0pt]{\rotatebox[origin=c]{90}{
	\centering
	\begin{tabular}{crclclclclcl}
		\hline \hline \vspace{-0.8\normalbaselineskip} \\
		\multirow{2}{*}{ Size } & \multirow{2}{*}{ Region } & & 20 keV & & 30 keV & & 50 keV & & 100 keV & & 1 MeV \\ \cline{4-4} \cline{6-6} \cline{8-8} \cline{10-10} \cline{12-12} \vspace{-0.8\normalbaselineskip} \\
		& & & $\mu$Gy/hist (\%$\Delta$) & & $\mu$Gy/hist (\%$\Delta$) & & $\mu$Gy/hist (\%$\Delta$) & & $\mu$Gy/hist (\%$\Delta$) & & mGy/hist (\%$\Delta$) \\ \hline \vspace{-0.8\normalbaselineskip} \\
		\multirow{3}{*}{ small } & nucleus & & 204.79 (0.01\%) & & 307.44 (0.09\%) & & 512.21 (0.06\%) & & 1023.38 (0.05\%) & & 10.243 (0.05\%) \\
		& cytoplasm & & 204.83 (0.03\%) & & 307.39 (0.08\%) & & 512.20 (0.05\%) & & 1024.37 (0.05\%) & & 10.233 (0.06\%) \\ \cline{4-4} \cline{6-6} \cline{8-8} \cline{10-10} \cline{12-12} \vspace{-0.8\normalbaselineskip} \\
		& theoretical & & 204.77 & & 307.16 & & 511.93 & & 1023.85 & & 10.239 \\ \vspace{-0.4\normalbaselineskip} \\
		\multirow{3}{*}{ medium } & nucleus & & 204.28 (0.01\%) & & 306.51 (0.04\%) & & 510.74 (0.02\%) & & 1021.57 (0.03\%) & & 10.216 (0.03\%) \\
		& cytoplasm & & 204.35 (0.05\%) & & 306.56 (0.06\%) & & 510.85 (0.04\%) & & 1021.67 (0.04\%) & & 10.205 (0.08\%) \\ \cline{4-4} \cline{6-6} \cline{8-8} \cline{10-10} \cline{12-12} \vspace{-0.8\normalbaselineskip} \\
		& theoretical & & 204.25 & & 306.38 & & 510.63 & & 1021.26 & & 10.213 \\ \vspace{-0.4\normalbaselineskip} \\
		\multirow{3}{*}{ large } & nucleus & & 203.26 (0.03\%) & & 304.94 (0.05\%) & & 508.19 (0.04\%) & & 1016.48 (0.05\%) & & 10.166 (0.06\%) \\
		& cytoplasm & & 203.36 (0.08\%) & & 305.00 (0.07\%) & & 508.31 (0.06\%) & & 1016.57 (0.06\%) & & 10.169 (0.09\%) \\ \cline{4-4} \cline{6-6} \cline{8-8} \cline{10-10} \cline{12-12} \vspace{-0.8\normalbaselineskip} \\
		& theoretical & & 203.20 & & 304.79 & & 507.99 & & 1015.98 & & 10.160 \\ \hline \hline \vspace{-0.8\normalbaselineskip} \\
	\end{tabular}}}
\end{table}

\begin{table}[htbp]
	\centering
	\vspace{0.5\normalbaselineskip}
	\captionl{Efficiency}{The efficiency ratios of the single cell Fano simulations over the recursive boundary simulations for the three cell sizes with medium nucleus sizes.  Estimates of efficiency are based on calculation times on a single thread AMD Ryzen 7 3800X (3.9 GHz) core.  Efficiency is computed as $\frac{1}{s^2t}$ where $s$ is nucleus uncertainty and $t$ is simulation time.
	\label{tab:Eff}}
	\small
	\begin{tabular}{rccc}
		\hline \hline \vspace{-0.8\normalbaselineskip} \\
		\multirow{2}{*}{\makecell{energy\\keV}} & \multicolumn{3}{c}{\makecell{relative efficiency (lattice/single)}} \\ \cline{2-4} \vspace{-0.8\normalbaselineskip} \\
		& small cell & med. cell & large cell \\ \hline \vspace{-0.8\normalbaselineskip} \\
		20   & 2.6$\times10^0$ & 2.6$\times10^0$ & 4.4$\times10^0$ \\ 
		30   & 1.3$\times10^1$ & 1.3$\times10^1$ & 1.5$\times10^1$ \\
		50   & 1.2$\times10^2$ & 1.2$\times10^2$ & 1.2$\times10^2$ \\
		100  & 1.7$\times10^2$ & 1.6$\times10^2$ & 1.5$\times10^2$ \\
		1000 & 8.7$\times10^5$ & 1.6$\times10^6$ & 2.5$\times10^6$ \\ \hline \hline \vspace{-0.8\normalbaselineskip} \\
	\end{tabular}
	\vspace{0.5\normalbaselineskip}
\end{table}
\newpage

\section{Discussion} \label{sec_EGSnrc:Discussion}

This work introduces the egs\_lattice geometry library, which allows for both modeling of (potentially) infinite numbers of geometries and establishing recursive boundary conditions (for Fano tests and other applications), and is used throughout much of the following validation work.  EGSnrc has passed the electron Fano cavity test to a 0.1\% threshold when simulating transport through a cell geometry with nanostructures.  EGSnrc is demonstrated to simulate transport through nanometer geometries and score dose in microcavities as reliably as other MC codes often used for microdosimetry.  The egs\_lattice library is currently available as a pull request on the EGSnrc GitHub distribution ``develop'' branch, but is planned to be incorporated into the master distribution in the future.

EGSnrc simulations were shown to be comparable to other, independent MC codes.  S-value calculations performed in EGSnrc with electron cutoff energy of 1~keV (42~\nm range in water\cite{Se16a}) have good agreement with the calculations of Sefl \etal\cite{Se15b} (Geant4-DNA, 7.4~eV cutoff) for source energies ranging 1-100~keV.  The agreement between the ratios of dose-to-tissue containing GNPs over dose to the homogeneous tissue/gold blend calculated in EGSnrc (Table~\ref{tab:Koger}) and those calculated by Koger and Kirkby\cite{KK16} in PENELOPE (electron cutoff energy of 100~eV) demonstrates EGSnrc's (and egs\_lattice's) ability to model nanostructures and further validates its ability to score dose in microscopic cavities.

Beyond the cross-validation, the electron Fano cavity test on a cell containing GNPs shows transport consistency results with EGSnrc when tested against analytical dose calculations.  The Fano test demonstrates two things when simulating GNPs in EGSnrc, (1) the 1~keV cutoff does not impact measured cellular dose values which are within 0.1\% of their expected value even if the cell contains GNPs, and (2) EGSnrc's electron transport algorithm maintains ideal CPE conditions even when transporting through thousands of GNPs with chord lengths of only a few nanometers.

The electron Fano cavity test\cite{Bo15} is chosen for this work over a more traditional approach of a regenerative photon simulation\cite{Ka99b} for efficiency gains.  The cell geometries tested are very small and complex (relative to a simple cavity) and the $<$0.01\% uncertainties on calculations are very computationally intensive.  The electron Fano cavity test allows for the use of the recursive boundary conditions in EGSnrc, which (with the geometry specifications in Section~\ref{ssec:Methods-Fano}) ensure that 3-27\% of the total simulation volume is occupied by the cell; this fraction could be further increased for more efficiency for each different cell size (if required).  The electron Fano cavity test results agree with the 0.1\% threshold found for EGSnrc using the more traditional (regenerative photon) Fano test on ionization chambers\cite{Ka99b,Se02} (independent of electron transport step size), even lower than the 0.3\% threshold in Geant4\cite{Po05} (step size parameters: \textit{dRoverRange}=0.01, \textit{finalRange}=0.1) and 0.2-0.4\% threshold for PENELOPE\cite{SA06} (step parameters: \textit{C1}=\textit{C2}=0.02, \textit{WCC}=0.002E$_0$  and \textit{WCR}=0.0002*E$_0$ for initial energy E$_0$).  A recent study by Lee \etal\cite{Le18} using the electron Fano cavity test on ionization chambers for the purposes of magnetic field testing found that the electron Fano cavity test passed to the same thresholds as above for Geant4, PENELOPE, and EGSnrc as the regenerative photon tests when the magnetic field was set to zero.

Both EGSnrc and the new egs\_lattice geometry operate reliably when scoring in \ums{}-scale targets, which, with their combined efficiency, open up new avenues of research.  Previously, a prototype of the egs\_lattice geometry was used to simulate over 6$\times10^{15}$~GNPs within a $\sim$25~\cms{}$^3$ volume\cite{MT17}.  This geometry would have taken over eight million terabytes of memory to model each GNP conventionally, which would have been unworkable on the 4~GB of memory per core available.  Even within this work, egs\_lattice has allowed for modeling of an infinite number of cells (each of which also contain tens of thousands of GNPs) to achieve Fano test results up to 2.52$\times10^6$ times faster than alternate methods, which allowed for Fano test simulations at energies as high as 1~MeV; these tests would have been computationally infeasible without recursive boundary conditions.  The new egs\_lattice geometry could also be employed in new fields, beyond GNPT, such as modeling biological micro-structures (\eg cells, trabeculae, and alveoli) on large scales, or modeling large detectors with many repeated elements very efficiently.

\section{Conclusion} \label{sec_EGSnrc:Conclusion}

A new lattice geometry library, egs\_lattice, was introduced into the egspp geometry package of EGSnrc.  The lattice geometry, as well as EGSnrc in general, underwent testing in this work to demonstrate reliability within microscopic geometries.  EGSnrc results scoring cell dose and dose ratios to microcavities containing GNPs agreed with independent calculations in Geant4-DNA\cite{Se15b} and PENELOPE\cite{KK16}, respectively.  EGSnrc with the added lattice geometries also passes the Fano cavity test (sub-0.1\% error) for a complex geometry comprised of a cell and GNPs, a test made possible using the recursive boundary conditions established using a cubic lattice to achieve efficiency gains of a factor of over a million in some cases.  Thus, EGSnrc can reliably calculate cell energy deposition and simulate radiation transport through GNPs.  This reliability, along with much of EGSnrc's inherit efficiency when modeling accurate electron transport\cite{AM15}, makes EGSnrc an excellent choice for investigations on short length scales and opens the door for further EGSnrc applications.

\section*{Acknowledgments}

The authors acknowledge support from the Kiwanis Club of Ottawa Medical Foundation and Dr. Marwah, the Natural Sciences and Engineering Research Council of Canada (NSERC), Canada Research Chairs (CRC) program, an Early Researcher Award from the Ministry of Research and Innovation of Ontario, and the Carleton University Research Office, as well as access to computing resources from Compute/Calcul Canada and the Shared Hierarchical Academic Research Computing Network (SHARCNET).  The authors declare that they do not have any conflict of interest.

\setlength{\baselineskip}{0.55cm}
%

\end{document}